\documentclass[twocolumn,prl,showpacs,superscriptaddress]{revtex4}
\usepackage{graphicx}
\usepackage{latexsym}
\usepackage{amsmath,calc}

\newcommand{\sx}[1]{{\scriptstyle #1}}

\begin{document}

\title
{A Hermite-Pad{\'e} perspective on Gell-Mann--Low renormalization group: an application to the correlation function of Lieb-Liniger gas}
\author{Vanja Dunjko}
\email{vanja.dunjko@umb.edu}
\affiliation{Department of Physics, University of
Massachusetts Boston, Boston, Ma 02125, USA}
\author{Maxim Olshanii}
\affiliation{Department of Physics, University of
Massachusetts Boston, Boston, Ma 02125, USA}
\date{\today }

%
\begin{abstract}
While Pad{\'e} approximation is a general method for improving convergence of series expansions, Gell-Mann--Low renormalization group normally relies on the presence of special symmetries. We show that in the single-variable case, the latter becomes an integral Hermite-Pad{\'e} approximation, needing no special symmetries. It is especially useful for interpolating between expansions for small values of a variable and a scaling law of known exponent for large values. As an example, we extract the scaling-law prefactor for the one-body density matrix of the Lieb-Liniger gas. Using a new result for the 4th-order term in the short-distance expansion, we find a remarkable agreement with known ab initio numerical results.
\end{abstract}

\maketitle
Quantitative information very often comes in the form of a series expansion in a dimensionless variable \cite{Beyond_Perturbation}. Often one knows only the first few terms; but even if one knows many, the raw series is usually useful only for a limited range of values of the expansion parameter. One may therefore turn to various methods for improving the convergence properties. This is justified if one expects that the solution, in the region of interest of the expansion variable, is an analytic continuation of the solution in the region in which one is performing the expansion. The best-known method is the Pad{\'e} approximation \cite{Pade_book} and its various Hermite-Pad{\'e} generalizations \cite{main_Hermite-Pade}.

On the other hand, when extrapolations from one parameter region to another were called for within particle physics, the solution was the Gell-Mann--Low renormalization group (GML-RG) \cite{GLM_RG_history_short,Weinberg}. This method is normally not comparable to the Pad{\'e} approximation because it takes as input not only the known terms of the expansion, but also certain known nontrivial symmetries in the problem, an approach that was later formalized and extended to the study of differential equations using Lie group-theoretic methods \cite{RG_in_differential_equations}. We should also mention a related body of work in Ref. \cite{Yukalov}.

We analyze the mechanism of operations of GML-RG in the single-variable (SV) case. We show that the SV GML-RG requires no special symmetry and is in fact an integral Hermite-Pad{\'e} approximation \cite{Hunter_Baker}, a connection that remains largely unexplored. The method is applicable to any series expansion, but is especially good for interpolation between a power-series expansion (for small values of a variable) and a scaling law (for large values of the variable), where the scaling law exponent is a known continuous function of a parameter, and one seeks, e.g., an approximation for the prefactor. As a case in point, we treat the density matrix of the Lieb-Liniger gas \cite{Lieb-Liniger}, an object of
considerable interest to both cold gases \cite{Olshanii03,Astrakharchik} and integrable systems \cite{Korepin_book} communities.

The GML-RG suggests itself very naturally in problems where a physical quantity at one value of a parameter is given as an expansion in terms of the value of the same quantity at another value of the parameter, a situation usually caused by regularization and renormalization \cite{Weinberg}. A simple single-parameter example is
$
T(k_{b}) = T(k_{a})
     + 2\mu \ln (k_{b}/k_{a})
         T^{2}(k_{a})
      +  (2\mu)^{2} \ln^{2} (k_{b}/k_{a})
            T^{3}(k_{a})+ \cdots\,,
$
which is the Born series for the scattering problem for the 2D delta-function potential \cite{RG_in_QM}; $T(k)$ is the complex-valued function one is trying to determine ($T$-matrix), $\mu>0$ is a constant (mass), and the expansion holds for all $k_{a},k_{b}>0$ (momenta). The expansion would not seem to be useful unless $k_{b}/k_{a} \approx 1$; however, we may try a ``divide and conquer'' approach: instead of going from $k_{a}$ to $k_{b}$ in a single large step, we use many smaller steps, $k_{a}=k_{0}\to k_{1} \to k_{2} \to \cdots \to k_{N}=k_{b}$, for each of which $k_{j+1}/k_{j} \approx 1.$ If $k_{n+1}/k_{n}=1+\epsilon$, we may let $\epsilon \to 0$, $N \to \infty$ so that $(1+\epsilon)^{N}=\textrm{const.}=k_{b}/k_{a}$. We obtain an equation of the form $k\,T'(k)=\beta(T(k))$, the Callan-Symanzik equation (CSE) for the problem, whose solution is the RG-improved solution of the problem. In this example, $\beta(T)=2\mu\, T^{2}$ at all Born orders. Thus, starting with the first nontrivial Born order, the RG treatment here produces the \emph{exact} result, $1/T(k_{b}) = 1/T(k_{a})-2\mu \ln (k_{b}/k_{a})$. Now note that the CSE may be obtained by applying $\left.\partial/\partial k_{b}\right|_{k_{b}, k_{a}\to k}$ to both sides of the expansion. Thus, to get the RG-improved solution, we differentiate the original expansion and then promptly integrate it, which raises the question: why do we not end up right back where we started? More formally, let $T(k)$ be the quantity of interest; then for some $r$, $\tau=k^{r}T$ is dimensionless. The exact solution of the problem may be written as $\tau (k_{b})=H(\tau (k_{a}),\, k_{a},\, k_{b})$ for some function $H$. Dimensional analysis then says that there must exist a function $G$ such that $\tau (k_{b})=G \left[\tau (k_{a}),\, k_{b}/k_{a}\right]$ (\textit{scale invariance} or \textit{self-similarity} of
$\tau$). Applying $\left.\partial/\partial k_{b}\right|_{k_{b}, k_{a}\to k}$, we get the CSE, $k\,\tau' (k)=\beta (\tau (k))$, where $\beta(x)=G^{(0,\,1)}(x,1)$. This is easily integrated, $\frac{k_{b}}{k_{a}}=\exp{\int_{\tau (k_{a})}^{\tau
(k_{b})}\,\frac{1}{\beta (t)}\,dt}$, and we see that if we start with the exact dependence of $\tau (k_{b})$ on $\tau (k_{a})$ and $k_{b}/k_{a}$ (i.e. exact $G$), the RG procedure must give the same exact dependence back. On the other hand, if we start with a truncated series expansion of $G$ (a finite-degree polynomial in $\tau (k_{a})$), then the RG does \emph{not} give the same finite-degree polynomial back. But how does the RG ``know'' that the polynomial was not itself the exact $G$? The answer is that not every function of two variables can play the role of exact $G$, because an exact $G$ must satisfy a special \emph{group property} \cite{GLM_RG_history_short}. Namely, if we wish to relate $\tau (k_{b})$ and $\tau (k_{a})$, we can do
it directly,
$\tau (k_{b})=G(\tau (k_{a}),\, k_{b}/k_{a}),$
or we can go through an intermediary, $\tau
(k_{a}) \to \tau (k_{c}) $ and then $\tau (k_{c}) \to \tau (k_{b})$. The results will agree only if $G$ satisfies \mbox{$
G(x,ab)=G(G(x,a),b)
$}
for all $a,$ $b$, and $x$.  This is a group property, because it allows us to introduce a (Lie) group with elements
$\mathcal{G}_{a}$ by defining group action on numbers through
$\mathcal{G}_{a}x=G(x,a).$ Group
multiplication then corresponds to composition,
$\mathcal{G}_{b}\mathcal{G}_{a}x=G(G(x,a),b)=G(x,ab)=\mathcal{G}_{ab}x$.  The generator gives the $\beta$-function: $\mathcal{B}x=\frac{d}{da}\left.\mathcal{G}_{a}\right|_{a=1}x=\beta(x)$.

The analysis that follows is straightforward, though it seems that it has not yet been reported in the literature. First, if $G(x,z)$ is to satisfy the group property, it cannot be a polynomial of finite degree in $x$, because then $G(G(x,a),b)$ would be a polynomial whose order is the square of that of $G(x,b a)$. Second, for any function of two variables $F(x,z)$ (regardless of whether is satisfies the group property or not) we may define $\beta(x)=F^{(0,\,1)}(x,1)$. The solution of the CSE is a function $\widetilde{F}(x,z)$ which satisfies the group property, has the same $\beta$-function as $F(x,z)$, and is equal to $F(x,z)$ if and only if $F(x,z)$ itself satisfies the group property. Next, suppose $F(x,z)$ is a truncated expansion in $x$ of some true $G(x,z)$, so that it does not satisfy the group property; the key question is why $\widetilde{\tau}(x)=\widetilde{F}(\tau_{0},x/x_{0})$  is so frequently a better approximation of $\tau_{\text{true}}(x)=G(\tau_{0},x/x_{0})$ than $\tau(x)=F(\tau_{0},x/x_{0})$ is. At first glance, the answer may seem obvious: because $\tau_{\text{true}}(x)$ is expressed in a self-similar form (through $G$), its approximation should also be represented in a self-similar form; $\widetilde{\tau}(x)$ is so represented (through $\widetilde{F}$), while $\tau(x)$ is not.
However, this cannot be right, because \emph{all} functions of single variable are representable in a self-similar form, at least on intervals where they are monotonic: $f(x_{2})=f(\frac{x2}{x1}x1)=f({\scriptstyle \frac{x2}{x1}}f^{-1}({\scriptstyle f(x_{1})}))$. (Provably, a function $G$ satisfies the group property if and only if there is a single-variable function $f$ such that $G(x,z)=f(z\,f^{-1}(x))$.) When this procedure is applied to $\tau(x)=F(\tau_{0},x/x_{0})$, we may call it ``an upgrade to a self-similar form through an \emph{exact inversion} of the approximate expression.'' However, instead of exactly inverting the polynomial that is $F(\tau_{0},x/x_{0})$, we may also think of it as a truncation of an infinite series and perform an order-by-order series reversion; then we compose the truncated series and its reversion according to $f(z\,f^{-1}(x))$. This way, any expansion in a single variable can be represented as a truncated self-similar expansion, which is then treatable by the GML RG (see below for a simpler method). However, the question of why $\widetilde{\tau}(x)$ is ``better'' than $\tau(x)$ remains.
To proceed, let $G$ be a function satisfying the group property; let
$z=x_{2}/x_{1}$ so that $\tau(x_{2})=G(\tau(x_{1}),\,z)$ and
assume $G$ has an expansion in powers of $\tau_{1}=\tau(x_{1})$:
$
G(\tau_{1},\,z)=f_{1}(z)\,\tau_{1}+f_{2}(z)\,\tau^{2}_{1}+f_{3}(z)\,\tau^{3}_{1}+\cdots
$. Imposing the group property and equating terms of like
orders in $\tau_{1}$, we obtain a recursive sequence of functional
equations:
$f_{1}(ba) = f_{1}(b)f_{1}(a)$;
$f_{2}(ba) = f_{1}^2(b)f_{2}(a)+f_{2}(b)f_{1}(a)$;
$f_{3}(ba) = f_{1}^3(b)f_{3}(a)+2f_{1}(b)f_{2}(b)f_{2}(a)+
        f_{3}(b)f_{1}(a)$; $\ldots\,$
We get $f_{1}(z)=z^{r}$, where $r$ is a real number, and
$f_{2}(z)=a_{2}\,f_{2,2}$, $f_{3}(z)=a_{2}^{2}\,f_{3,2}+a_{3}\,f_{3,3}$, etc., where $f_{2,2}=\frac{z^{r}}{r}(z^{r}-1)$, $f_{3,2}=\frac{z^{r}}{r^{2}}(z^{r}-1)^{2}$, etc. (If $r=0$, we have $f_{2,2}=\ln z$; $f_{3,2}=\ln^{2} z$; $f_{3,3}=\ln z$, etc.)
Here $a_{2}={f_{2}}'(1)$, $a_{3}={f_{3}}'(1)$, etc., are arbitrary constants.  It will be convenient to introduce an expansion in ``rows:''\\
\vspace{-1.5\baselineskip}
\begin{tabbing}
\= $\sx{\tau_{2}}$ \; \=$\sx{=}$
\=\mbox{}\;$\sx{z^{r}\tau_{1}}$\;\;\;\;\;\;\=$\sx{+}$
 \=$\sx{a_{3}\,f_{4,3}\,\tau_{1}^{4}}$ \=$\sx{+}$
 \=$\sx{a_{3}a_{2}^{2}\,f_{4,3}\,\tau_{1}^{4}}$ \=$\sx{+}$
 \=$\sx{a_{3}a_{2}^{2}\,f_{4,3}\,\tau_{1}^{4}}$ \=$\sx{+}$\kill
\>$\sx{\tau_{2}}$ \>$\sx{=}$
\>$\sx{f_{1}(z)\,\tau_{1}}$ \>$\sx{+}$
\>\;\;$\sx{f_{2}(z)\,\tau^{2}_{1}}$ \>$\sx{+}$
\>\;\;$\sx{f_{3}(z)\,\tau^{3}_{1}}$ \>$\sx{+}$
\>\;\;$\sx{f_{4}(z)\,\tau^{4}_{1}}$ \>$\sx{+}$\; $\sx{\cdots}$\\
\>\>\>\rule[0.5\baselineskip]{
\widthof{
 \mbox{}\;$\sx{z^{r}\tau_{1}}$\;\;\;\;\;\;$\sx{+}$
 $\sx{a_{3}\,f_{4,3}\,\tau_{1}^{4}}$ $\sx{+}$
$\sx{a_{3}a_{2}^{2}\,f_{4,3}\,\tau_{1}^{4}}$ $\sx{+}$
 $\sx{a_{3}a_{2}^{2}\,f_{4,3}\,\tau_{1}^{4}}$ $\sx{+...}$
}
}{0.1ex}
\end{tabbing}
\vspace{-2.0\baselineskip}
\begin{tabbing}
\= $\sx{\tau_{2}}$ \; \=$\sx{=}$
\=\mbox{}\;$\sx{z^{r}\tau_{1}}$\;\;\;\;\;\;\=$\sx{+}$
 \=$\sx{a_{3}\,f_{4,3}\,\tau_{1}^{4}}$ \=$\sx{+}$
 \=$\sx{a_{3}a_{2}^{2}\,f_{4,3}\,\tau_{1}^{4}}$ \=$\sx{+}$
 \=$\sx{a_{3}a_{2}^{2}\,f_{4,3}\,\tau_{1}^{4}}$ \=$\sx{+}$\kill
\> \>$\sx{=}$
\>\;\;\; $\sx{z^{r}\tau_{1}}$ \>$\sx{+}$
\>\quad\; $\sx{0}$ \>$\sx{+}$
\>\quad\; $\sx{0}$ \>$\sx{+}$
\>\quad\; $\sx{0}$ \>$\sx{+}$\; $\sx{\cdots}$\\
\> \>
\> \>$\sx{+}$
\>$\sx{a_{2}\,f_{2,2}\,\tau_{1}^{2}}$ \>$\sx{+}$
\>\; $\sx{a_{2}^{2} \,f_{3,2}\,\tau_{1}^{3}}$ \>$\sx{+}$
\>\; $\sx{a_{2}^{3}\, f_{4,2}\,\tau_{1}^{4}}$ \>$\sx{+}$\; $\sx{\cdots}$\\
\> \>
\> \>
\> \>$\sx{+}$
\>\; $\sx{a_{3}\,f_{3,3}\,\tau_{1}^{3}}$ \>$\sx{+}$
\> $\sx{a_{3}a_{2}^{2}\,f_{4,3}\,\tau_{1}^{4}}$ \>$\sx{+}$\; $\sx{\cdots}$\\
\> \>
\> \>
\> \>
\> \>$\sx{+}$
\> $\sx{a_{4}\,f_{4,4}\,\tau_{1}^{4}}$ \>$\sx{+}$\; $\sx{\cdots .}$
\end{tabbing}
\vspace{-0.5\baselineskip}
Note that $f_{n}(z)\,\tau_{1}^{n}$ is the sum of the entries of the $n$th column; the $n$th order of perturbation theory gives the first $n$ columns. Suppose we knew the expansion to the second order. We see that if $f_{2}(z)$ is nonzero, then $a_{2}$ is nonzero, and so even though we do not know $f_{3}(z)$, we can nevertheless conclude that it must at least have the term proportional to $a_{2}^{2}$. In fact the entire second row in the table must be nonzero. (In the 2D scattering example above, all the other rows were zero.) More generally, from the first $n$ columns we can deduce the full first $n$ rows. What RG does, in effect, is fill out all the rows whose entries can be deduced from the known columns, sets the other rows to 0, sums the result to infinite order, and analytically continues beyond the radius of convergence of the resulting infinite series. To see this, note that the $\beta$-function is determined solely by the leading terms of each row: $\beta(\tau_{1})=r\,\tau_{1}+a_{2}\,\tau_{1}^{2}+a_{3}\,\tau_{1}^{3}+\cdots$. The solution of the CSE with the $\beta$-function truncated to the $n$th order then is precisely an $\widetilde{F}(\tau,\,z)$ whose ``row expansion'' has just the first $n$ rows. In short, the RG treatment results in a \emph{minimal upgrade} to a self-similar form: it only includes the rows we know how to reconstruct from the perturbation expansion. In contrast, an upgrade through an \emph{exact inversion} of the approximate expression will in addition produce spurious rows, artifacts that have nothing to do with the underlying problem. Note that when we use an ordinary perturbation expansion to compute actual numerical values, we set to zero all orders we did not actually compute; in RG we simply follow the same rule, though this time we apply it to the rows.
\newcommand{\derv}[3]{\text{T}[#1\,|\,\mbox{\scriptsize $\text{T}_{#2}\{#3 \}$}]}

There is also a completely different way to think about SV GML-RG, which makes it clear that it is a species of Hermite-Pad{\'e} approximation. From the representation $G(x,z)=f[z\,f^{-1}(x)]$ it is easy to show that $\beta\{f\}(x)=f^{-1}(x)/[f^{-1}]'(x)$. Now consider integrals of the form $\exp{\int_{x_{a}}^{x_{b}}\,1/g(t)\,dt}$. If $g(t)=\beta\{f\}(x)$, the integral is identically equal to $f^{-1}(x_{b})/f^{-1}(x_{a})$, and represents the integrated CSE. We are interested, however, in the situation when we only know a truncated expansion for $f$, $T\{\mbox{\footnotesize $f$}\}$. By formal series manipulation, we can compute two objects that are asymptotically (for small $x$) equal to $\beta\{f\}$: (a) $\beta\{T\{\mbox{\footnotesize $f$}\}\}$, and (b) $T\{\mbox{\footnotesize $\beta\{f\}$}\}$. If $g$ is set to (a), namely to $T\{\mbox{\footnotesize $f^{-1}$}\}/T\{\mbox{\footnotesize $[f^{-1}]'$}\}$, the integral $\exp{\int_{x_{a}}^{x_{b}}\,1/g(t)\,dt}$ will return $T\{\mbox{\footnotesize $f^{-1}$}\}(x_{b})/T\{\mbox{\footnotesize $f^{-1}$}\}(x_{a})$. Setting $x_{a}=0$ and performing a series reversion then brings us right back to where we started, to $T\{\mbox{\footnotesize $f$}\}$, as it must because every step was reversible. However, if $g$ is set to (b), namely  to $T\{\mbox{\footnotesize $f^{-1}/[f^{-1}]'$}\}$, that is an irreversible step, and the result is something new: an \textit{RG approximant}. More formally: suppose $f$ has an expansion (for simplicity about zero) $c_{0}x^{\alpha_{0}}+c_{1}x^{\alpha_{1}}+\cdots$ with all $c_{j}\neq 0$ and $\mathsf{Re}\,\alpha_{0}<\mathsf{Re}\,\alpha_{1}<\cdots$. Let $\text{T}_{n}\{f\}(x)$ be the truncation of that expansion up to and including the $x^{\alpha_{n}}$ term. Now formal series manipulations allow one to compute truncated expansions of various transformations of $f$ (such as $f^{2}$, $1/f$, $f'$, $f^{-1}$, etc.) order-by-order, up to some maximal order that is still completely determined by $\text{T}_{n}\{f\}(x)$. Let us denote these derived expansions by, e.g., $\derv{f^2}{n}{f}(x)$ (for example, $\derv{1/\sin}{2}{\sin}(x)=\frac{1}{x}+\frac{1}{6}\,x$).
Now we
introduce the (order-$\alpha_{n}$, centered-at-$0$) \textit{renormalization
group on-the-inverse approximant (RGIA)} through
\begin{multline}
\text{RGIA}_{n}\{f\}(x) =\\ \lim_{\epsilon \to 0} \,
\text{T}_{n}\{f\}(\epsilon) \, \exp
\int_{\epsilon}^{x}\,\frac{1}{\derv{f/f'}{n}{f}(s)}\,ds\,
. \label{RGIA}
\end{multline}
This expression is the key result of the paper. Note that: (1) the usual RG result is given by $\text{RGIA}_{n}\{f^{-1}\}(x)$; (2) usually $\derv{f/f'}{n}{f}(s)$ is a polynomial, in which case the integral can be evaluated analytically (in terms of logarithms and arctangents) once the polynomial is decomposed into partial fractions. This requires factoring the polynomial, in other words, finding its zeros, which is normally the only numerical step in the computation. Replacing the \emph{denominator} in Eq. (\ref{RGIA}) by its Pad{\'e} approximation tends to improve results further.
\newcommand{\betai}{\beta_{\text{I}}}

There is at least one class of functions $f$ for which it is obvious that $\text{RGIA}_{n}\{f\}$ must be a better approximant for $f$ than $\text{T}_{n}\{f\}$: the functions whose only singularities are poles (of any order) but whose derivatives have no zeros. The reason is that for such functions, $f/f'$ is analytic everywhere. An example is $f(x)=\tan x$, with $f/f'=\frac{1}{2}\sin 2x$. Near the pole at $x=\pi/2$, the Taylor expansion for $f$ fails as an approximation for $f$. However, the Taylor expansion for $f/f'$ remains a good approximation for $f/f'$, and so $\text{RGIA}_{n}\{f\}$ remains a good approximation for $f$. Order-by-order, in this case, at least, it also performs better than the Pad{\'e} approximation. One should emphasize that experience thus far shows that the RGIA is useful for a much broader class of functions, despite the fact that there is no clear reason why it should be so; but in the case of Pad{\'e} approximation, it is also still not completely understood why \emph{it} performs as well as it does, even after more than a century of research \cite{Pade_book}.

It is however clear that RGIA is a special case of the so-called integral Hermite-Pad{\'e} approximant \cite{Hunter_Baker}, defined as follows: one finds the polynomials $P(x)$, $Q(x)$, and $R(x)$, of orders $p$, $q$, and $r$, that satisfy the condition $P(x)f'(x)+Q(x)f(x)+R(x)=\mathcal{O}(x^{p+q+r+2})$, with $P(0)=1$; the more terms one has in the expansion of $f$, the higher the order of agreement, $p+q+r+2$, that one can demand. Then one sets the right hand side to zero; the solution of the resulting equation is the integral approximant. We see that RGIA corresponds to $R(x)=0$ and $Q(x)=-f'(0)/f(0)$. Thus, the SV GML-RG is an integral Hermite-Pad{\'e} approximation plus a series reversion \cite{Baik95}: $P(x)[f^{-1}]'(x)-\frac{[f^{-1}]'(0)}{f^{-1}(0)}f^{-1}(x)=\mathcal{O}(x^{p+2})$. We note in passing that while in the Pad{\'e} approximation community it is believed that best approximants should be invariant under the homographic transformation $x=Aw/(1+Bw)$ \cite{Baker_Ising}, this invariance demands $p=q+2$ (and $q=r$), so that the RGIA is invariant under this transformation only if $p=2$. On the other hand, high-energy physics community has accumulated a huge amount of knowledge and experience about GML-RG. In the future we plan to bring the two research traditions together around the point of contact, the RGIA, and begin to work out what mutual implications they may have.

Turning to practical applications of the RGIA, we will consider problems where, in addition to the expansion of $f(x)$ for small values of $x$, one also knows the exponent of the scaling law for large values of $x$: $f(x) \sim C/x^{\alpha}$ with $\alpha$ known but $C$ unknown. The idea is to apply the RGIA to $\widetilde{f}(x)=x^{\alpha}f(x)$ (one can treat the case where $f \sim C (x_{0}-x)^{\beta}$ at a finite $x_{0}$ in a similar way). Problems like this can be approached using Pad{\'e} approximation as well \cite{Baker_Ising,Bernu01}, but for exponents which are not simple fractions, the Pad{\'e} method becomes inconvenient. For example, to treat an irrational exponent, one would have to approximate it by a ratio of integers $p/q$, and then find the approximant for $[\widetilde{f}(x)]^{q}$. This becomes particularly unnatural if the exponent is a function of a continuous parameter. In this case RGIA is clearly a better choice, because even for irrational exponents $\alpha$, $\beta=\widetilde{f}/\widetilde{f}'$ is a rational function; the RGIA will then itself be a continuous function of the exponent. As a nontrivial application, we consider the one-body--reduced density matrix $\rho_{1}(z_{1};\,z'_{1}) = \int_{0}^{L}\ldots\int_{0}^{L}\!\!\!\,\,\, dz_{2}\,\ldots dz_{N}
\Psi(z_{1},\,z_{2},\,\ldots,\,z_{N})
\Psi^{*}(z'_{1},\,z_{2},\,\ldots,\,z_{N})$ of the ground-state Lieb-Liniger gas (a 1D gas of bosons interacting via interparticle potential $g_{\text{1D}}\,\delta(z)$) with periodic boundary conditions \cite{Lieb-Liniger,Korepin_book,Olshanii03,Astrakharchik}. In the thermodynamic limit, where the number of particles $N$ and the system length $L$ go to infinity but the 1D density  $n=N/L$ is kept constant, $L\rho_{1}(0;,\,z)=g_{1}(z)/n$ is finite and its short-distance expansion, $n z \to 0$, is given as $g_{1}(z)/n =
1+c_{2}(\gamma)\left(n z\right)^{2}+c_{3}(\gamma)\left|n
z\right|^{3}+c_{4}(\gamma)\left(n z\right)^{4}+\cdots\,.$ Here $\gamma=(m/\hbar^{2})\,g_{\text{1D}}/n$ is the dimensionless strength of the interaction.  In \cite{Olshanii03} we showed that $c_{1}(\gamma)=0$, $c_{2}(\gamma)=-\frac{1}{2}\left[e_{2}(\gamma)-\gamma
e'_{2}(\gamma)\right]$, and  $c_{3}(\gamma)=\frac{1}{12}\gamma^{2}e'_{2}(\gamma)$, where $e_{m}(\gamma)=\left[\gamma/\lambda(\gamma)\right]^{m+1}
\!\!\int_{-1}^{1}\,x^{m}\,g\left(x|\gamma\right)\,dx$ is the normalized $m$th moment of the ``density of rapidities'' function $g\left(x|\gamma\right)$, which, together with the function $\lambda(\gamma)$, is determined through the Lieb-Liniger system of equations: $\mbox{$2\pi\,g\left(x|\gamma\right)$}-\mbox{$\int_{-1}^{1}\,
[2\lambda(\gamma)]/[\lambda^{2}(\gamma)+(y-x)^{2}]\,g\left(y|\gamma\right)
\, dy$}=1$ and
$\gamma \int_{-1}^{1}\, g\left(x|\gamma\right) \, dx \,
=\lambda(\gamma)$. We also use a new result,
$
c_{4}(\gamma)=\gamma\, e'_{4}/12-3\, e_{4}/8
+\left(2\,\gamma^{2}+\gamma^{3}\right)\,e'_{2}/24
-\gamma\, e_{2}/6-\gamma\, e_{2}\,e'_{2}/4+3\,e_{2}^{2}/4\,,
$
whose derivation will appear elsewhere \cite{Olshanii09}. For \mbox{$n z \to \infty$}, from Luttinger liquid theory \cite{Haldane} it is known that \mbox{$g_{1}(z)/n = C(\gamma)/\left(n z\right)^{\alpha(\gamma)}$} with
$8\pi^{2}\alpha^{2} =
              6e_{2}(\gamma)
             -4\gamma e_{2}'
             +\gamma^2e_{2}''
$; $C(\gamma)$ is \emph{not} known, and we extract it using GML-RG.
\begin{figure}
%
\includegraphics[scale=.5,draft=false]{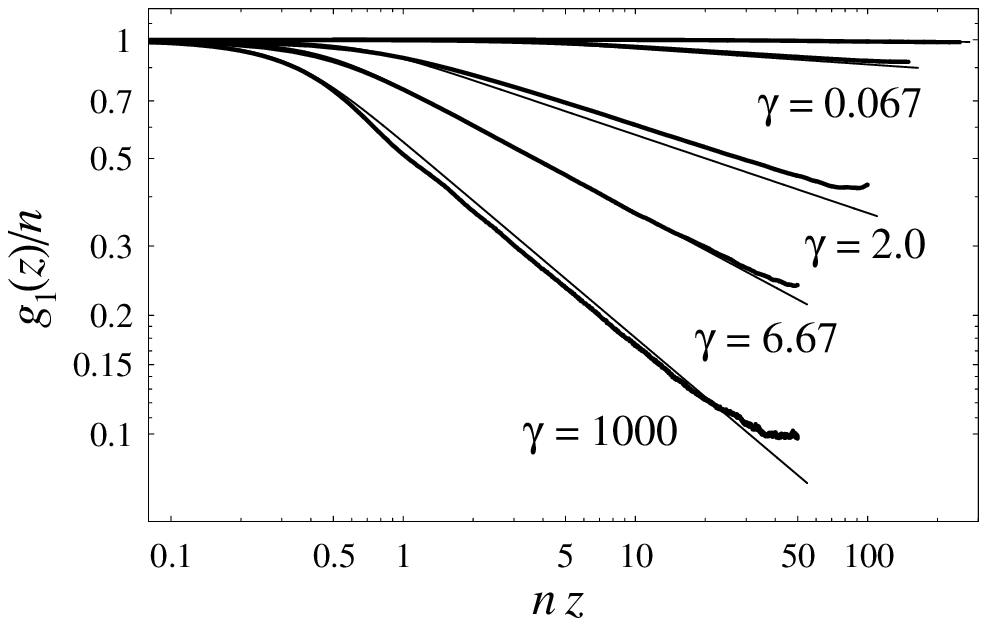}
\begin{minipage}[b]{\widthof{\includegraphics[scale=.5]{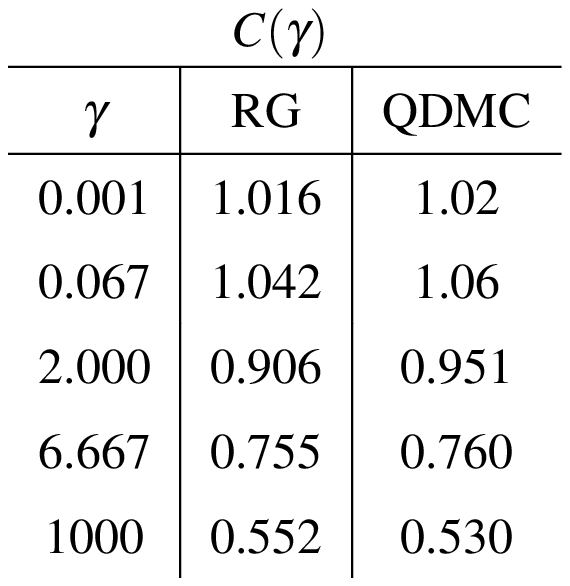}}}
\includegraphics[scale=.5,,draft=false]{Dunjko_Pade_RG_Fig1b.eps}
\vspace{0.0\baselineskip}
\mbox{}
\end{minipage}
\caption { \label{corr_func} One-body--reduced density matrix of the Lieb-Liniger gas for a variety of interaction strengths $\gamma$. Thick lines: results of an ab initio quantum diffusion Monte Carlo (QDMC) simulation; thin lines: an RGIA interpolation between the fourth-order short-distance expansion and the long-distance scaling law. Not labeled are the nearly horizontal curves for $\gamma=0.001$. To the right of the plot is a table comparing the values for the prefactor of the long-distance scaling law obtained from the RGIA interpolation, and from the QDMC.
         }
\end{figure}
In Figure \ref{corr_func}, we compare the results of an RGIA interpolation between the short-distance expansion and the long-distance scaling law for $g_{1}(z)/n$ with the results of an ab initio quantum diffusion Monte Carlo simulation \cite{Astrakharchik}, at a wide range of interaction strengths $\gamma$. In the table to the right of the plot, we compare the results for the prefactor $C(\gamma)$, finding remarkable agreement. (We should mention here the excellent analytical estimate of $C(\gamma)$ from an inspired, though as-yet unjustified, extrapolation of Popov theory in Ref. \cite{Cazalilla}.)

We have analyzed the mechanism of operation of single-variable Gell-Mann--Low renormalization group (SV GML-RG). We found that it does not rely on any special symmetries and that it is in fact is a combination of series reversion and an integral Hermite-Pad{\'e} approximation. To the best of our knowledge, such an explicit connection between the GML-RG and Pad{\' e} methods had not yet been pointed out, and future work may use this connection to explore the implications that the GML-RG and Pad{\'e} research traditions have for each other. One obvious direction would be to see if invariance under homographic transformation plays any role in GML-RG.
Finally, we singled out a class of interpolation problems for which GML-RG is particularly well-suited; we demonstrated the effectiveness of the method in the case of the density matrix of the Lieb-Liniger gas, for which we have also announced the exact form of the fourth-order term in the short-distance expansion.

We thank R. Shakeshaft, M. Jaffrey, and M. Rigol for fruitful discussions, and G. E. Astrakharchik for providing us the data from his QDMC calculations. This work was supported by the Office of Naval Research grant \mbox{No. N00014-06-1-0455} and the National Science Foundation grant \mbox{No. PHY-0754942}.

%


\end{document}